# The sensitivity of refractive index sensors based on defected 1D photonic crystals: an analytical approach

*Nikolay A. Vanyushkin\*, I.M. Efimov, Ashot H. Gevorgyan*

Far Eastern Federal University, 10 Ajax Bay, Russky Island, Vladivostok, 690922, Russia
\*E-mail: vaniuschkin.nick@ya.ru

In this paper, an analytical formula for the sensitivity of optical sensors based on one-dimensional photonic crystals (PCs) with a defect was derived for the first time. Based on this formula, a comparative analysis of the sensitivity of defected PCs with and without mirror symmetry was carried out. In addition, the exact values of sensitivity in the limit of an infinite PC with a quarter-wave unit cell were obtained. It was shown that the sensitivity of a defective PC with an infinitely thick defect layer is equal to that of a perfectly reflecting Fabry-Perot resonator and does not depend on the specific structure of the PC. The results of this work provide a significant simplification of the analysis and optimization of optical sensors based on defective PCs, as well as a better understanding of the numerous numerical results obtained previously.

## 1. Introduction

With the development of optics, optical sensors have been actively used in science and industry. An optical sensor is a device that uses the optical properties of an object to measure various parameters. Nowadays, optical sensors are already widely used in various fields, including industry, robotics, medicine and even household devices. Optical sensors work based on the use of light; they can be active, generating their own light and then measuring its reflection, transmission through the object or absorption by the object, or they can be passive, registering only light from other sources [1-4]. One of the promising areas of development of this direction is the creation of sensors based on 1D photonic crystals (PCs) with a defect layer (DL) in the structure [5-10]. PCs are optical media that have a periodic or quasi-periodic change in refractive index. The main peculiarity of PCs is the presence of the so-called photonic bandgap (PBG) in their transmission spectrum. The incident radiation with wavelengths falling into the PBG cannot propagate through the PC and is reflected from it. If a defective layer is added to the periodic structure of the PC, the periodicity of the structure is broken, which leads to a change in the transmission and reflection spectra in the whole region. This manifests itself in the appearance of a narrow bandwidth inside the PBG, which is called a defect mode (DM). The position of the DM depends on the parameters of the DL, such as the thickness and refractive index. This fact allows us to find the refractive index of the DL by measuring the wavelength of the DM. Among the main advantages of PC-based sensors are high sensitivity, ease of design and fabrication, and compactness. Over the last decades, many different optical sensors based on PCs with DL have been developed for a wide range of applications. Among them are gas sensors [11-13], liquid sensors [14], temperature sensors [15], biochemical sensors [16,17]. Usually, when developing sensors based on defective PCs, numerical calculations are used to find the DM wavelength and other parameters. However, there are a number of works in which analytical expressions for the DM wavelength have been obtained [18-22]. Usually, it is done through the condition of mode localization inside the resonator (field decay when moving away from the DL), or through the condition that the reflection coefficient of the whole structure inside the PBG is equal to zero.

In addition, one of the important parameters of the sensor is its sensitivity [23,24], which denotes how large the DM shift at the refractive index change. The research showed the advantage of defected PCs with mirror symmetry (MS) which have a higher sensitivity, especially when the DL borders with high refractive index layers. Nevertheless, to the best of our knowledge, analytical formulas for the sensitivity of sensors based on defective PCs have not been obtained anywhere. This paper is devoted to the solution of this problem.

## 2. Theory

Let us consider the reflection and transmission of a plane electromagnetic wave through a 1D defected binary PC (Figure 1) which consists of the DL sandwiched between two PCs, which are made of $N$ unit cells each and playing the role of mirrors, which together with the DL form a Fabry-Perot resonator. The unit cell consists of the layer A with refractive index $n_1$ and thickness $d_1$ and the layer B with refractive index $n_2$ and thickness $d_2$. For clarity we will assume that $n_2 > n_1$ throughout the paper. The exact layer order in the unit cell can be ether AB or BA and be the same or reversed for the two mirrors. The regions outside the structure are filled with vacuum ($n_0 = 1$).

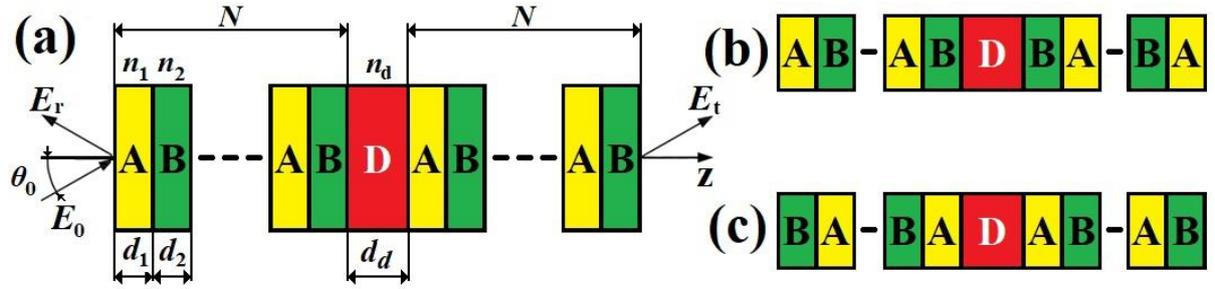

**Figure 1.** (a) Geometry of the problem using the example of a structure without MS (AB|D|AB). Arrangement of the layers relative to the DS (b) with the MS and a higher refractive index at the DL boundary (AB|D|BA), (c) with the MS and a lower refractive index at the DL boundary (BA|D|AB).

When solving this problem, we employed the transfer matrix approach to calculate the transmission spectra of the PC [25]. The transfer matrix relates the field amplitudes at different interfaces and can be presented for *j*-th layer in the following form:

$$M_j = \begin{pmatrix} \cos k_j d_j & -\dfrac{i}{p_j} \sin k_j d_j \\ -i p_j \sin k_j d_j & \cos k_j d_j \end{pmatrix}, \quad (1)$$

where $k_j = (\omega/c) n_j \cos\theta_j$, $d_j$ is the thickness of the layer, $\theta_j$ is the angle of refraction which is determined by $n_j \sin\theta_j = n_0 \sin\theta_0$. Also $p_j = n_j \cos\theta_j$ for *s*-wave and $p_j = (1/n_j)\cos\theta_j$ for *p*-wave. Then the transfer matrix *m* of a PC consisting of $N$ layers can be obtained by multiplying the transfer matrices of all layers:

$$m = M_1 M_2 ... M_{N-1} M_N = \begin{pmatrix} m_{11} & m_{12} \\ m_{21} & m_{22} \end{pmatrix}. \quad (2)$$

Carrying out some mathematics then the complex transmission and reflection coefficients for *s*- and *p*-waves are given as

$$t_N = \frac{2 p_0}{(m_{11} + m_{12} p_0) p_0 + (m_{21} + m_{22} p_0)}, \quad (3)$$

$$r_N = \frac{(m_{11} + m_{12} p_0) p_0 - (m_{21} + m_{22} p_0)}{(m_{11} + m_{12} p_0) p_0 + (m_{21} + m_{22} p_0)}. \tag{4}$$

Here $p_0 = n_0 \cos\theta_0$ for s-wave and $p_0 = (1/n_0)\cos\theta_0$ for p-wave.

On the other hand, the transfer matrix of any structure without absorption or gain can be expressed through the reflection $r$ and transmission $t$ coefficients of that structure as follows

$$M = \begin{pmatrix} \dfrac{1}{t} & \dfrac{r^*}{t^*} \\ \dfrac{r}{t} & \dfrac{1}{t^*} \end{pmatrix}. \tag{5}$$

The DL transfer matrix without taking into account the reflection from the boundaries will then take the following form

$$M_d = \begin{pmatrix} e^{-i\varphi} & 0 \\ 0 & e^{i\varphi} \end{pmatrix} \tag{6}$$

where

$$\varphi = \frac{2\pi}{\lambda} n_d h_d \tag{7}$$

is the change of phase of the wave at a single passage through the DL.

The transfer matrix of a defective PC can be generally represented as

$$m = (M_{01} M_1 M_{1d}) M_d (M_{d2} M_2 M_{20}) = M_I M_d M_{II} \tag{8}$$

where $M_1$, $M_d$, $M_2$ are "inner" transfer matrices of PC$_1$, DL and PC$_2$, i.e. these transfer matrices relate the fields at inner sides of their outer boundaries. $M_{01}$ and $M_{20}$ are "boundary" matrices for the interfaces between the perfect PCs and the external environment, i.e. these transfer matrices relate the fields at different sides of the same boundary. Similarly, the matrices for the boundaries between the perfect PCs and DL are defined as $M_{1d}$ and $M_{d2}$. For convenience, these transfer matrices can be combined into both left $M_I$ and right $M_{II}$ mirror' matrices.

If we write equation (8) in explicit form through reflection and transmission coefficients, we obtain

$$\begin{pmatrix} \dfrac{1}{T} & \dfrac{R^*}{T^*} \\ \dfrac{R}{T} & \dfrac{1}{T^*} \end{pmatrix} = \begin{pmatrix} \dfrac{1}{t_I} & \dfrac{r_I^*}{t_I^*} \\ \dfrac{r_I}{t_I} & \dfrac{1}{t_I^*} \end{pmatrix} \begin{pmatrix} e^{-i\varphi} & 0 \\ 0 & e^{i\varphi} \end{pmatrix} \begin{pmatrix} \dfrac{1}{t_{II}} & \dfrac{r_{II}^*}{t_{II}^*} \\ \dfrac{r_{II}}{t_{II}} & \dfrac{1}{t_{II}^*} \end{pmatrix}. \tag{9}$$

Here $R$ and $T$ are the reflection and transmission coefficients of the whole structure:

$$R = \frac{e^{2i\varphi} t_I r_{II} + t_I^* r_I}{e^{2i\varphi} t_I r_I^* r_{II} + t_I^*}, \tag{10}$$

$$T = \frac{e^{i\varphi} t_I t_I^* t_{II}}{e^{2i\varphi} t_I r_I^* r_{II} + t_I^*}. \tag{11}$$

Since the reflection coefficient $R$ is zero for DMs [26], then by demanding $R = 0$ in (10) we obtain the characteristic equation for the DM

$$e^{2i\varphi} t_I r_{II} + t_I^* r_I = 0 \tag{12}$$

which we can further simplify by expanding the reflection and transmission coefficients into amplitude and phase

$$r = |r| e^{i\rho}, \quad t = |t| e^{i\tau}. \tag{13}$$

Substituting (13) into (12) we obtain the necessary phase $\varphi_d$ condition for DM

$$\varphi_d = \frac{1}{2}(\rho_I - 2\tau_I - \rho_{II} - i \ln\left|\frac{r_I}{r_{II}}\right|) + \pi q = -\frac{1}{2}(\tilde{\rho}_I + \rho_{II} + i \ln\left|\frac{\tilde{r}_I}{r_{II}}\right|) + \pi q = \varphi_0 + \pi q, \quad q \in \Box. \tag{14}$$

Here we introduced $\tilde{r}_I = -r_I^* t_I^* / t_I = |\tilde{r}_I| e^{i\tilde{\rho}_I}$ (i.e. $\tilde{\rho}_I = -\rho_I - 2\tau_I + \pi - i \ln|r_I/\tilde{r}_I|$) [27] which is the reflection coefficient of the left mirror when the wave is incident from the right, i.e. from the DL side. The solution (14) for the DL phase is periodical with the period $\pi$, and the integer number $q$ can be associated with the DM order [23].

It is worth to pay special attention to the meaning of the phase $\varphi_d$ from (14), as well as its difference from $\varphi$ introduced earlier in (7). On the one hand, the phase $\varphi$ simply shows the change of the phase of the wave when passing the DL and depends solely on the properties of the DL and the wavelength of light, but only at some values of $\varphi$ a DM is present. Usually, we can control $\varphi$ by changing the thickness of the DL $d_d$ and choosing a different wavelength $\lambda$. On the other hand, $\varphi_d$ shows us what $\varphi$ should be equal to achieve zero reflection coefficient at a given wavelength and with given mirrors' parameters. Thus, $\varphi$ represents the actual phase of the DL, while $\varphi_d$ is the required phase of the DL at which a DM will be observed.

Figure 2 shows the phase $\varphi_d$ spectra from equation (14) for the first (without MS) and the second (with MS) structure, as well as the reflection spectrum $|R|^2$ at $d_d = 0$, that is, when $\varphi = 0$. It is well seen that all wavelengths with $|R|^2 = 0$ correspond to the condition $\varphi_d = \varphi = 0$.

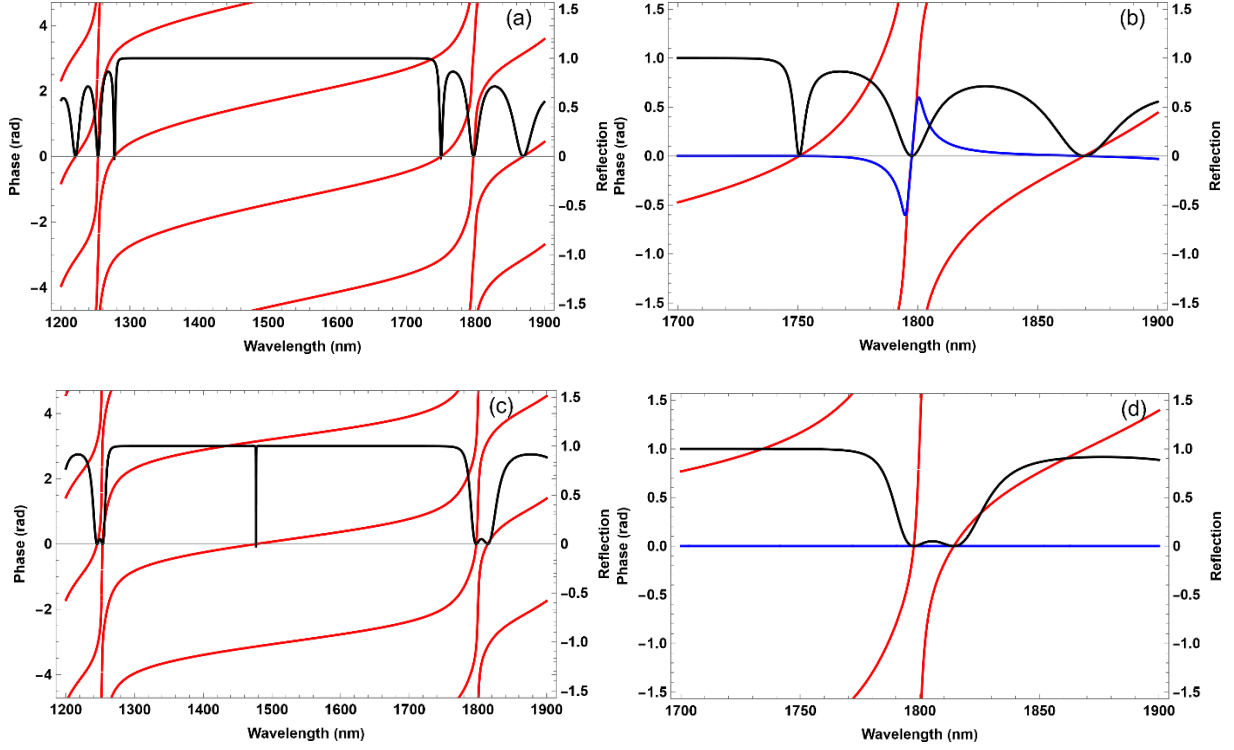

**Figure 2.** Spectra of the real (red line) and imaginary (blue line) part of $\varphi_d$ and reflection $|R|^2$ (black line) for (a,b) the first structure without MS and (c,d) for the second structure with MS. Parameters of the structures:

$$N = 10, \; n_1 = 1.5, \; n_2 = 2.4, \; d_d = 0, \; n_d = 1.3, \; \Lambda = d_1 + d_2 = 1000 \text{ nm}, \; n_1 d_1 = n_2 d_2.$$

If we analyze expression (14), we can understand the reasons for various phenomena associated with defective PCs. For example, the difference in the absolute values of the reflection coefficients of two PCs leads to the appearance of a nonzero imaginary part of $\varphi_d$ (see Figure 2(b)). This is due to the fact that in a structure without mirror symmetry the reflection coefficients from the defect boundaries are different, which leads to a difference in the reflection coefficients of the two mirrors, and in particular, their absolute values. Since for the reflection coefficient of the whole structure to be zero, the condition $\varphi = \varphi_d$ must be exactly satisfied, this leads to the necessity of an imaginary component of the refractive index. In the case of the real $n_d$ we have $\text{Im}\,\varphi \neq \text{Im}\,\varphi_d$ and the reflection coefficient at the DM wavelength will be different from zero, i.e. the DM amplitude decreases. The inequality of reflection coefficients of two PCs can be associated, in particular, with a different number of unit cells or if the DL borders with layers with different refractive index. For this reason, the use of a defective PC with MS guarantees the maximum DM amplitude in the absence of absorption [23].

**2.1 Refractive index sensitivity**

One of the most important sensor parameters is sensitivity. In the case of refractive index sensors, this value usually indicates how much the characteristic wavelength (operating wavelength) $\lambda$ shifts at the same change in the refractive index $n$ being measured, and can be defined as

$$S = \frac{1}{\lambda} \frac{d\lambda}{dn}. \tag{15}$$

In the case of sensors based on defective PCs $\lambda$ is the wavelength of DM $\lambda_d$, and $n$ is the refractive index $n_d$ of DL. Further, to obtain an analytical formula for the sensitivity of our sensor, it is sufficient to differentiate the equation $\varphi = \frac{2\pi}{\lambda_d} n_d h_d = \varphi_d$ by refractive index $n_d$ taking into account that $\varphi_d = \varphi_d(n_d, \lambda_d)$ and $\lambda_d = \lambda_d(n_d)$:

$$\frac{d\varphi_d}{dn_d} = \frac{\partial \varphi_d}{\partial n_d} + \frac{\partial \varphi_d}{\partial \lambda_d} \frac{d\lambda_d}{dn_d}, \tag{16}$$

$$\frac{d\varphi}{dn_d} = 2\pi h_d \left( \frac{1}{\lambda_d} - \frac{n_d}{\lambda_d^2} \frac{d\lambda_d}{dn_d} \right). \tag{17}$$

By combining (16) and (17) we obtain the expression for sensitivity

$$S = \frac{1}{\lambda_d} \frac{d\lambda_d}{dn_d} = \frac{1}{n_d} \frac{\varphi_d - n_d \frac{\partial \varphi_d}{\partial n_d}}{\varphi_d + \lambda_d \frac{\partial \varphi_d}{\partial \lambda_d}}. \tag{18}$$

Since the function $\varphi = \varphi_d = \varphi_0 + \pi q$ grows indefinitely with increasing DL thickness $d_d$, which is accompanied by an increase in the order of DM $q$, while the derivatives $\frac{\partial \varphi_d}{\partial n_d} = \frac{\partial \varphi_0}{\partial n_d}$ and $\frac{\partial \varphi_d}{\partial \lambda_d} = \frac{\partial \varphi_0}{\partial \lambda_d}$ are bounded and equal for all orders of DM $q$, then we can draw an important conclusion, that $\lim_{d_d \to \infty} S = 1/n_d$ regardless of the specific properties of the two PCs. This result is not accidental and coincides with the expression for the sensitivity of the sensor based on a Fabry-Perot resonator with perfectly reflecting mirrors, for which the condition for eigenmodes $\lambda = \frac{nL}{2q}$ is satisfied, where $L$ is the resonator length, $q$ is integer number, and $n$ is refractive index of the medium inside the resonator. Thus, at large DL thickness, as one would expect, the defective PC starts to behave like a conventional Fabry-Perot resonator, which imposes a fundamental limitation on the maximum sensitivity.

**3. Numerical results**

In this section, we present the sensitivity dependences using specific structures as examples and compare them. Figure 1 shows three variants of the arrangement of PC layers with respect to DL: (a) periodic PC without MS with respect to DL, (b) PC with MS, where DL borders with layers with a higher refractive index, and (c) PC with MS, where DL borders with layers with a lower refractive index.

Figure 3(a,b,c) shows the phase $\varphi_d$ spectra (using the zeroth mode with $q = 0$ as an example) and its derivatives $n_d \frac{\partial \varphi_d}{\partial n_d}$ and $\lambda_d \frac{\partial \varphi_d}{\partial \lambda_d}$, as well as the reflection coefficient $|R|^2$, within the first PBG for the three structures. The higher-order DM phase $\varphi_d$ has the same dependence but is shifted by $\pi q$. In this case, as discussed earlier, the derivatives are the same for all orders of modes. Figures 3(d,e,f) show the sensitivity spectra of several orders of DM within the PBG for the three structures.

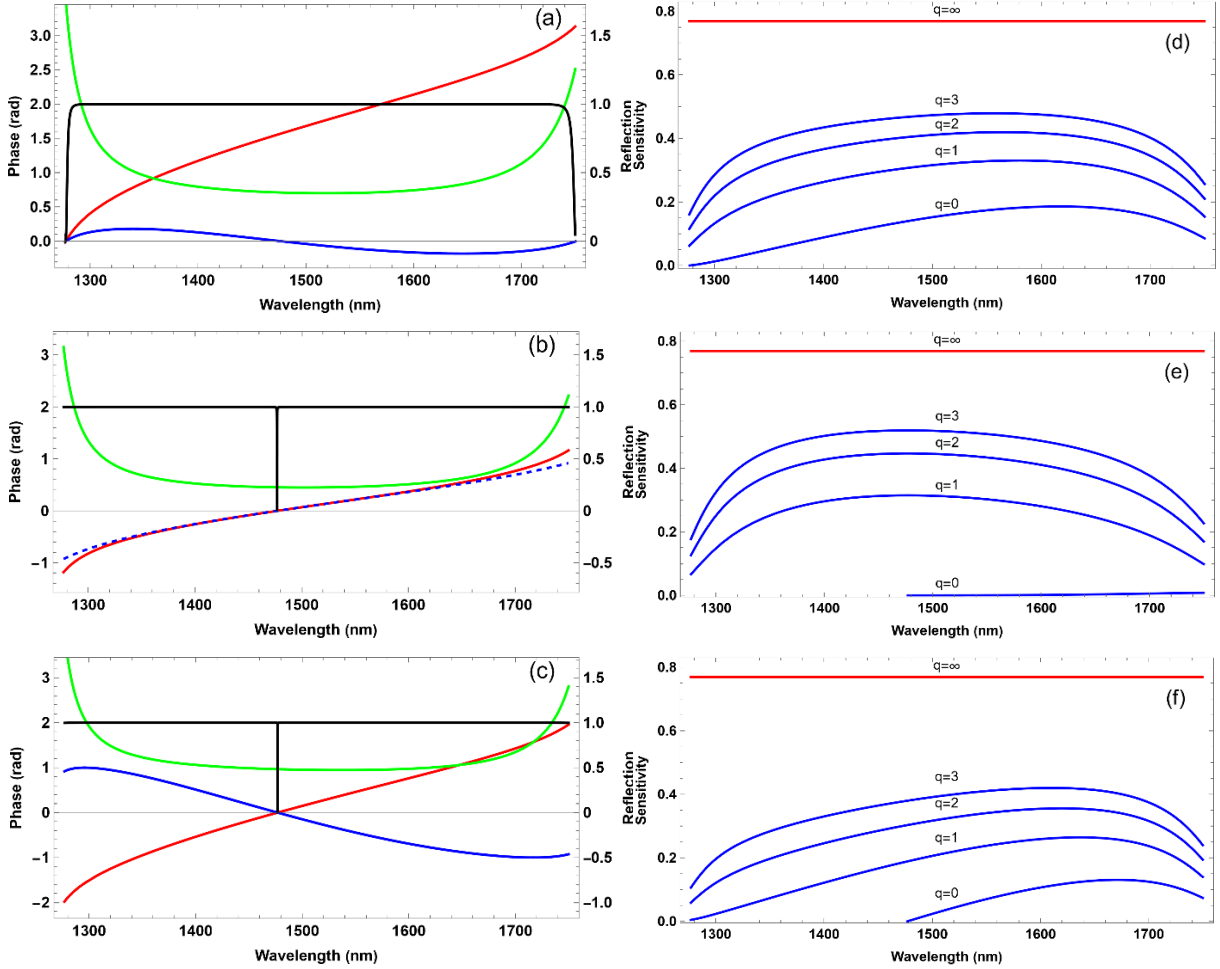

**Figure 3**. (a,b,c) The spectra of phase $\varphi_d$ (red line) for zeroth order DM ($q = 0$) and its derivatives $n_d \frac{\partial \varphi_d}{\partial n_d}$ (blue line) and $0.1 \times \lambda_d \frac{\partial \varphi_d}{\partial \lambda_d}$ (green line), as well as the reflection $|R|^2$ (black line). (d,e,f) Sensitivity spectra of several orders of DMs. The red horizontal line shows the sensitivity $S_\infty = 1/n_d$ of the infinite order DM (infinitely thick DL). Parameters are the same as in Figure 2.

Figure 3(a) shows that in the structure without MS, at zero thickness of the DL ($\varphi_d = 0$), the DM is located at the short-wavelength edge of the PBG and moves to the long-wavelength region as $d_d$ increases. On the other hand, in the structures with MS, the DM (Figures 3(b,c)) always starts moving inside the PBG, and if the unit cell is a quarter-wave stack, the starting point coincides with the center of the PBG. Further, a comparison of the spectra of all terms in formula (18) shows that in all structures the derivative $\lambda_d \frac{\partial \varphi_d}{\partial \lambda_d}$ in the denominator has the greatest influence on the sensitivity of DM of small orders because it is an order of magnitude larger than the phase $\varphi_d$. Same derivative also leads to a sharp decrease in sensitivity at the edges of the PBG. On the other hand, the derivative $n_d \frac{\partial \varphi_d}{\partial n_d}$ is much smaller than the phase of the zeroth mode for the structure without MS and approximately equal to the phase in the case with MS, and therefore can have a significant influence on the zero mode at MS, but it is negligibly small for the first and large orders of DM in all structures. However, it is worth noting that it is $n_d \frac{\partial \varphi_d}{\partial n_d}$ that determines the wavelength with the greatest sensitivity. Thus in the second structure (Figure 2(e)) the maximum of sensitivity always coincides with the zero of this derivative, while in the other structures (Figure 2(d,f)) the maximum approximately coincides with the minimum of the derivative and shifts towards the center of the PBG as the DM order increases. In addition, it is worth noting that in the second structure the zero mode has extremely low sensitivity just from the fact that the difference $\varphi_d - n_d \frac{\partial \varphi_d}{\partial n_d}$ is practically zero within entire PBG.

**4. Quarter-stack PC sensitivity**

It should be noted that in the PC with a quarter-wave unit cell, for which $n_1 d_1 = n_2 d_2$ is true, it is possible to obtain an exact value for the reflection coefficient $R$ [25], which allows to considerably simplify the expression (18). For this purpose, let us consider a defective PC whose DM is exactly in the center of the first PBG, i.e. $\lambda_d = 2(n_1 d_1 + n_2 d_2)$. Phases of complex reflection coefficients in this case are equal to $\rho_R = \tilde{\rho}_L = \pi$ or $\rho_R = \tilde{\rho}_L = 0$ in the case of MS and $\rho_R = \pi$, $\tilde{\rho}_L = 0$ or $\rho_R = 0$, $\tilde{\rho}_L = \pi$ in the case without MS. Thus, if we neglect the imaginary part of the phase, the condition on the DL phase in the center of the PBG becomes $\varphi_d = \pi q$ and $\varphi_d = \frac{\pi}{2} + \pi q$ for the structure with and without MS, respectively. This condition is satisfied at $d_d = 0$ in the case of MS, while without MS a minimum DL thickness $d_d = \lambda_d /(4 n_d)$ is required.

In the general case, the reflection coefficient of a PC [25] made of $N$ unit cells is represented as

$$r_N = \frac{C}{A - W_N}. \tag{19}$$

For a periodic PC with BA unit cell the coefficients in (19) have the form:

$$W_N = \frac{\sin(N-1)K\Lambda}{\sin NK\Lambda}, \tag{20}$$

$$\cos K\Lambda = \cos k_1 d_1 \cos k_2 d_2 - \frac{1}{2}\left(\frac{n_2}{n_1} + \frac{n_1}{n_2}\right) \sin k_1 d_1 \sin k_2 d_2, \tag{21}$$

$$A = e^{ik_1 d_1}(\cos k_2 d_2 - \frac{i}{2}(\frac{n_2}{n_1} + \frac{n_1}{n_2})\sin k_2 d_2), \tag{22}$$

$$C = e^{ik_1 d_1}\frac{i}{2}(\frac{n_2}{n_1} - \frac{n_1}{n_2})\sin k_2 d_2. \tag{23}$$

The coefficients $A$ and $C$ are diagonal elements of the transfer matrix of the unit cell, $K$ is Bloch wave vector and $\Lambda = d_1 + d_2$ is PC period. The corresponding expressions for the AB cell are easily obtained by swapping the layers' parameters.

Reflection coefficient of the wave incident from the DL to the PC is

$$r = \frac{r_N + r_d}{1 + r_d r_N}. \tag{24}$$

Here $r_d = \frac{n - n_d}{n + n_d}$, and $n$ is refractive index of the PC layer adjacent to the DL.

Next

$$\frac{\partial \varphi_d}{\partial \lambda_d} = -\frac{1}{2}(\frac{\partial \tilde{\rho}_I}{\partial \lambda_d} + \frac{\partial \rho_{II}}{\partial \lambda_d} + i\left|\frac{r_{II}}{\tilde{r}_I}\right|\frac{\partial}{\partial \lambda_d}\left|\frac{\tilde{r}_I}{r_{II}}\right|), \tag{25}$$

$$\frac{\partial \rho_{I,II}}{\partial \lambda_d} = \text{Im}(\frac{1}{r_{I,II}}\frac{\partial r_{I,II}}{\partial \lambda_d}), \tag{26}$$

$$\frac{\partial r}{\partial \lambda_d} = \frac{\partial r}{\partial r_N}\frac{\partial r_N}{\partial \lambda_d} = \frac{1 - r_d^2}{(1 + r_d r_N)^2}\frac{\partial r_N}{\partial \lambda_d}, \tag{27}$$

$$\frac{\partial r_N}{\partial \lambda_d} = \frac{1}{A - W_N}\frac{\partial C}{\partial \lambda_d} - \frac{C}{(A - W_N)^2}(\frac{\partial A}{\partial \lambda_d} - \frac{\partial W_N}{\partial \lambda_d}) = \frac{1}{A - W_N}\frac{\partial C}{\partial \lambda_d} - \frac{C}{(A - W_N)^2}\frac{\partial A}{\partial \lambda_d}. \tag{28}$$

In the last expression we used $\frac{\partial W_N}{\partial \lambda_d} = 0$, because $\frac{\partial W_N}{\partial \lambda_d} = \frac{\partial W_N}{\partial K}\frac{\partial K}{\partial \lambda_d}$ and $\frac{\partial K}{\partial \lambda_d} = 0$ at any $N$ in the center of PBG [25]. By using the equality of optical thicknesses in the expressions (19)-(23) we obtain for the BA unit cell

$$A = -\frac{n_1^2 + n_2^2}{2n_1 n_2}, \tag{29}$$

$$C = \frac{n_2^2 - n_1^2}{2n_1 n_2}, \tag{30}$$

$$\frac{\partial A}{\partial \lambda_d} = -i\frac{\pi}{\lambda_d}\frac{(n_1 + n_2)^2}{4n_1 n_2}, \tag{31}$$

$$\frac{\partial C}{\partial \lambda_d} = i\frac{\pi}{\lambda_d}\frac{n_2^2 - n_1^2}{4n_1 n_2}. \tag{32}$$

Further for simplicity we consider the case $N \gg 1$, for which $W_N \approx -\frac{n_1}{n_2}$ and $r \approx -1$ (BA cell) or $r \approx 1$ (AB cell). In addition, $|\tilde{r}_I| = |r_{II}| = 1$ within whole PBG for all structures. Finally, we obtain

$$\frac{\partial r_N}{\partial \lambda_d} = \pm i \frac{\pi}{\lambda_d} \frac{n_1}{n_2 - n_1}, \quad (33)$$

where the plus is for BA cell and minus for AB cell. The expression of $\frac{\partial r}{\partial r_N}$ in (27) differs depending on what layer is adjustment to the DL: $\frac{\partial r}{\partial r_N} = \frac{n_2}{n_d}$ for layer A and $\frac{\partial r}{\partial r_N} = \frac{n_d}{n_1}$ for layer B. Thus

$$\frac{\partial \rho_{AB}}{\partial \lambda_d} = \frac{\pi}{\lambda_d} \frac{1}{n_2 - n_1} \frac{n_1 n_2}{n_d}, \quad (34)$$

$$\frac{\partial \rho_{BA}}{\partial \lambda_d} = \frac{\pi}{\lambda_d} \frac{n_d}{n_2 - n_1}. \quad (35)$$

Then, the derivative $\frac{\partial \varphi_d}{\partial \lambda_d}$ for each structure is as follows:

for AB|D|AB

$$\lambda_d \frac{\partial \varphi_d}{\partial \lambda_d} = \frac{\pi}{2} \frac{1}{n_2 - n_1} (\frac{n_1 n_2}{n_d} + n_d), \quad (36)$$

for AB|D|BA

$$\lambda_d \frac{\partial \varphi_d}{\partial \lambda_d} = \pi \frac{n_d}{n_2 - n_1}, \quad (37)$$

for BA|D|AB

$$\lambda_d \frac{\partial \varphi_d}{\partial \lambda_d} = \pi \frac{1}{n_2 - n_1} \frac{n_1 n_2}{n_d}. \quad (38)$$

Following the same path, it is easy to obtain that for any $N$

$$n_d \frac{\partial \varphi_d}{\partial n_d} = 0 \quad (39)$$

in the center of the PBG for all structures, which can be seen, for example, in Figure 3(a,b,c). Thus from formula (39) it follows that sensitivity $S$ depends only on $\varphi_d$ and $\lambda_d \frac{\partial \varphi_d}{\partial \lambda_d}$. At $n_d^2 < n_1 n_2$ the sensitivity of the sensor in which the DL is adjacent to layers with a larger refractive index will be higher, which is consistent with the numerical results obtained in [23], but otherwise it is more advantageous to work with DL adjacent to layers with a lower refractive index. PCs without MS always occupy an intermediate position and are never optimal in terms of sensitivity.

It should be noted that although we obtained exact values of sensitivity in the approximation of infinite PCs, in practice the reflection coefficient in the center of the PBG grows very rapidly with increasing number of unit cells, and therefore these expressions are a good approximation for not too thin PCs. For example, for the PCs in Figure 3 with $N = 10$ the obtained formulas (36)-(38) have a relative error of the order of 0.01%.

## 5. Conclusions

To conclude, we have obtained for the first time an analytical formula for calculating the sensitivity of optical sensors based on 1D PCs with a defect in the structure. It is worth noting that this formula is not limited only to defective PCs, but is fully applicable to any Fabry-Perot type resonator for which the phase and amplitude of the reflection coefficients of the mirrors are known. It has been shown that there is a fundamental limitation on the sensitivity of the $S \leq 1/n_d$, which does not depend on the particular parameters of the mirrors. In addition, explicit expressions for the sensitivity for the quarter-wave PC at the center of the PBG in the infinite PC approximation were derived, which contain only the refractive indices of the unit cell and the refractive index of the DL. These expressions confirm the previously obtained conclusion about the advantage of the mirror symmetric structure, in which the DL borders with layers with a higher refractive index, but this is true only if the condition $n_d^2 < n_1 n_2$ is satisfied, and otherwise it is advantageous to swap the layers in the unit cell. The results of this work allow to considerably simplify the analysis and optimization of optical sensors based on defective PCs and allow to better understand numerous numerical results obtained earlier.

**Acknowledgements**
The work was supported by the Foundation for the Advancement of Theoretical Physics and Mathematics "BASIS" (Grant № 21-1-1-6-1).

**Conflict of interest**
No potential conflict of interest was reported by the authors.